\documentclass[twocolumn,showpacs,preprintnumbers,amsmath,amssymb]{revtex4}

\usepackage{graphicx}
\usepackage{dcolumn}
\usepackage{bm}
\usepackage{epsfig}

\begin{document}

\title{Magnetism of CoO polymorphs}

\author{Thomas Archer, Ruairi Hanafin and Stefano Sanvito}
\affiliation{School of Physics and CRANN, Trinity College, Dublin 2, Ireland}
\date{\today}

\begin{abstract}
A microscopic explanation for the room temperature ferromagnetism in diluted ZnO:Co is at present
rather elusive. Although standard secondary phases can usually be ruled out, it is less clear whether
regions with high Co concentration coexist with undoped portions of the film, i.e. whether 
some form of CoO polymorph can be responsible for the magnetic signal.
Since X-ray usually excludes the presence of the native rock-salt phase, the study of CoO polymorphs
becomes particularly interesting.
In this work we investigate theoretically the magnetism of CoO in both the wurtzite and
zincblende phases. By using a combination of density functional theory with the LDA+$U$ approximation
and Monte Carlo simulations we demonstrate that wurtzite and zincblende CoO have a complex
frustrated anti-ferromagnetic ground state with no net magnetic moment in the bulk. Most importantly
the estimated critical temperatures are well below room temperature for both cases. This suggests
that bulk CoO polymorphs are not responsible for the room temperature magnetism observed for ZnO:Co,
although the role of clusters with uncompensated spins or arranged in a spinodal decomposed structure 
still remains an open question.
\end{abstract}
\maketitle

\section{Introduction}
Diluted magnetic semiconductors (DMS)~\cite{DMS} are a new class of materials in which ordinary semiconductors 
are doped with transition metal ions, whose spins align in a ferromagnetic ground state. Their remarkable properties,
in particular the interplay between ferromagnetism and free carriers, promise a generation of novel electronic devices 
based on the spin degree of freedom \cite{spintronics}. Unfortunately after almost a decade of research the Curie 
temperature ($T_\mathrm{C}$) of GaAs:Mn, the most studied among all the DMS, is still at around 170~K \cite{nottingham} 
and it is not clear whether it will ever be possible to overcome all the limiting material issues \cite{DSM}. It is therefore 
understandable that the magnetic community became excited by the announcement of room temperature 
ferromagnetism in ZnO:Co \cite{ueda}.

ZnO is transparent, conducting \cite{TCO} and piezoelectric \cite{ZnOpiezo}. If ferromagnetism is also demonstrated, this 
will be the ultimate multi-functional material. Unfortunately, in contrast to GaAs:Mn, the phenomenology associated with 
ZnO:Co is extremely vast and often contradictory. Several models have been proposed to explain the experimentally observed 
room temperature ferromagnetism (RTF) including: the donor impurity band exchange  \cite{coey}, surface mediated 
magnetism \cite{surface}, Co/oxygen vacancies pairs magnetism \cite{das} and uncompensated antiferromagnetic 
nanoclusters \cite{DietlZnO}. 

In general it is often difficult to exclude the presence of secondary phases and indeed metallic Co clusters are often identified in thin 
films \cite{MikeSecond}. It is even more difficult to exclude the presence of high Co density regions. In these, the concentration 
of transition metals can exceed the percolation limit resulting in magnetism, as recently demonstrated for ZnTe:Cr \cite{DietlNM}. 
This result was then extrapolated to ZnO:Co and uncompensated spins at the surface of hypothetical CoO antiferromagnetic 
clusters were proposed as the source of the observed room temperature magnetism \cite{DietlZnO}.
Therefore, as wurzite (WZ) CoO can be considered the end member of the Zn$_{1-x}$Co$_x$O alloy, the
study of its magnetic properties becomes of paramount importance. In this work we investigate the magnetic
state of various CoO polymorphs, including rocksalt (RS), zincblende (ZB) and wurtzite (WZ), and conclude 
that these phases cannot support any room temperature magnetic order in the bulk. Our results relate directly
to the many bulk CoO polimorphs, which have been already experimentally synthesized but for which the magnetic
characterization is still scarce. They also exclude bulk  CoO as the source of magnetism in ZnO:Co, although the
clusters hypothesis with uncompensated spins at the surface of the clusters remains still an open question.

\section{Computational Details}
In this work we use a combination of density functional theory (DFT) and Monte Carlo (MC) simulations to
investigate both the ground state and the magnetic critical temperature $T_\mathrm{C}$, of CoO polymorphs.
Importantly we go beyond the simple local density approximation (LDA) and use the LDA+$U$ scheme in the
Czyzyk-Sawatzky form \cite{LDAU} as implemented \cite{gosia} in the pseudopotential code {\it Siesta} \cite{siesta}.
As a test of our scheme we have also carried out calculations with the rotationally invariant form
of the LDA+$U$ functional proposed in reference \cite{rot}. This yields only tiny differences in the total energy
differences for the cubic phase and it was not employed for the other polymorphs. The empirical Coulomb $U$ 
and exchange $J$ parameters are chosen to be $U$~=~5~eV and $J$~=~1~eV. These values reproduce the lattice 
constant of RS CoO in the ground state structure. Our $U$ and $J$ values are also in good agreement with previously 
determined values from constrained DFT~\cite{pickett}. In all our calculations we used norm-conserving Troullier-Martins' 
pseudopotentials \cite{troullier91} with non-linear core corrections \cite{zhu92} and a real-space regular grid with a grid 
spacing equivalent to a plane-wave cut-off of 800~Ry. Reciprocal space integration was performed on a grid with an 
equivalent real space distance of 20~\AA. We relaxed all structures until the forces and pressure are smaller than 
0.005~eV/\AA\ and 5~kbar respectively. The spin-orbit interaction is not included in our calculations, since it causes only tiny 
corrections to the total energy. The largest corrections are expected for the WZ structure for which electron paramagnetic
resonance (EPR) measurements of isolated Co ions within the ZnO lattice produce a zero-field split of
$D$~=~2.76~cm$^{-1}$ \cite{anisotropy}.

Supercells were constructed for the RS, WZ and ZB structures, containing 32, 48 and 36 atoms respectively. For each 
polymorph 62 total energy calculations were performed for randomly assigned collinear spin configurations. We then 
mapped the DFT energy onto the classical Heisenberg Hamiltonian
\begin{equation}
H_\mathrm{H}=E_0-\frac{1}{2}\sum_{i,j}J_{\vec{r_{ij}}}\vec{S}_i\cdot\vec{S}_j\;,
\end{equation}
where $J_{\vec{r_{ij}}}$ is the Heisenberg exchange constant, $\vec{S}_i$ the classical spin associated to the $i$-th site 
($|\vec{S}_i|=3/2$ for CoO) and $E_0$ the energy of the corresponding paramagnetic phase. We prefer the simple use of
collinear configurations over the more standard non-collinear scheme based on the magnetic force theorem \cite{bruno},
since we do not posses any {\it a priori} information on the nature of the ground state of the various polymorphs. We then 
estimate the error in our computed critical temperatures by comparing the calculated N\'eel temperature for RS
CoO with the experimental one and by transferring such an error to the other polymorphs. 

The effective Hamiltonian $H_\mathrm{H}$ was then used in our 
MC simulations to determine the ground state and $T_\mathrm{C}$. In the case of WZ CoO we also include
a uniaxial anisotropy term setting an hard-axis along the WZ $c$-axis, with the value for the zero-field split 
taken from EPR measurements $D$~=~2.76~cm$^{-1}$ \cite{anisotropy}. Thus the final Hamiltonian used for
the MC simulations is
\begin{equation}
H_\mathrm{H}=E_0-\frac{1}{2}\sum_{i,j}J_{\vec{r_{ij}}}\vec{S}_i\cdot\vec{S}_j\;+
\sum_{i}D(\vec{S}_i \cdot\hat{n})^2\:,
\end{equation}
where $\hat{n}$ is a unit vector along the WZ $c$-axis. Spins were reoriented using the standard Metropolis 
algorithm. The acceptance probability of a new state is 1 if the new configuration has a lower energy, otherwise it is given by 
the Boltzmann distribution $e^{- \Delta{E}/{k_{B}T}}$~\cite{Monte}, where $\Delta E$ is the energy difference between
the old and the new configurations. Each system was first equilibrated at a given temperature, then
the specific heat and Binder cumulants were calculated over several million MC steps.
These were used to extract $T_\mathrm{C}$. Simulations were performed with lattices containing
512 and 1000 Co atoms with periodic boundary conditions.

\section{Rock salt cobalt oxide}
\begin{figure}
\epsfxsize=8cm
\centerline{\epsffile{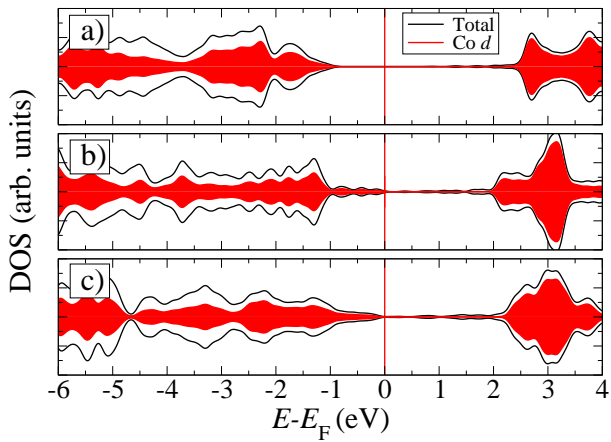}}
\caption{Projected density of states for (a) rock-salt, (b) wurtzite and (c) zinc-blende CoO.
The antiferromagnetic configurations used are type-II for the RS phase, $c$-type for WZ and
antiferromagnetic with alternating ferromagnetic planes along the [100] direction for the ZB.}
\label{Fig1}
\end{figure}

We begin our analysis by investigating RS CoO, since both its structure and magnetic properties are experimentally very well
established. This represents a good test for our computational scheme and it will also give us the opportunity 
of estimating its likely uncertainty. RS CoO is a type-II antiferromagnet (AFII) 
below the N\'eel temperature $T_\mathrm{N}~=~287$~K. In this magnetic configuration ferromagnetic planes align 
antiferromagnetically along the [111] direction~\cite{CoOtn}. Our calculated lattice parameters are reported in table \ref{Tab1} 
and agree by construction with previously published experimental data \cite{RS-CoO}.
\begin{table}
\begin{tabular}{lllllrrrrrr} \hline \hline
         & $a$         &    $c$         &   $u$  & V (\AA$^3$) & $E_0$ & $J_1$ & $J_2$ & $J_3$ & $J_4$ & $T_\mathrm{C}$ \\ \hline
RS         &    4.260         &     --            &  --   & 19.32 & 0 & 1.5 & -12.2 & -- & -- & 210 \\
WZ                                    &  3.244        & 5.203                 &  0.416   & 23.71   & 200.6 & 6.1 & -36.7 & -0.2 & -5.2  & 160   \\
WZ$^*$                            &  3.476        & 4.292                 &    0.500   & 22.05 & 120.6 & 0.0  & -55.2 & -0.8 & -24   & 100   \\
ZB                                         & 3.245         &     --               &   --  & 23.83  & 313.2 & -5.0 & 0.7 & 0.6 & -2.0 & 55 \\ \hline\hline
\end{tabular}
\caption{\label{Tab1}Summary of the calculated structural and magnetic properties for the various CoO
polymorphs: $a$, $c$ (in \AA) and $u$ (fractional) are the lattice constants, $V$ is the volume
per formula unit (in \AA$^3$), $E_0$ (in meV) is the Heisenberg energy of the paramagnetic phase, $J_n$
(in meV) are the exchange constants, and $T_\mathrm{C}$ (in K) is the critical temperature calculated
from the specific heat.
}
\end{table}
The calculated density of states (DOS) is shown in figure~\ref{Fig1}. The valence band is a hybrid band formed from
the O-$p$ and the Co-$d$ orbitals, while the conduction band is of purely $d$ in character. This places the material
on the Zaanen-Sawatzky-Allen~\cite{zaanen} diagram between charge transfer and Mott-Hubbard insulators, as
reported by several other calculations \cite{pickett}. The Mulliken populations for the Co-$d$ orbitals returns
a magnetic moment of 2.77$\mu_\mathrm{B}$ with no contributions from O, in good agreement
with the 2+ oxidation state.

\begin{figure}
\epsfxsize=8cm
\centerline{\epsffile{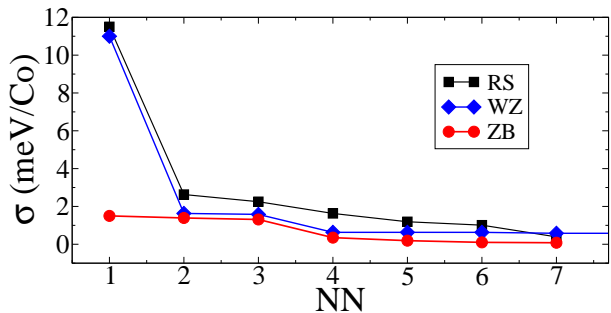}}
\caption{Standard deviation $\sigma$ for the fit of the calculated DFT total energies onto the
Heisenberg Hamiltonian as a function of the number of nearest neighbors (NN) included in the model.}
\label{Fig2}
\end{figure}

\begin{figure}
\epsfxsize=8cm
\centerline{\epsffile{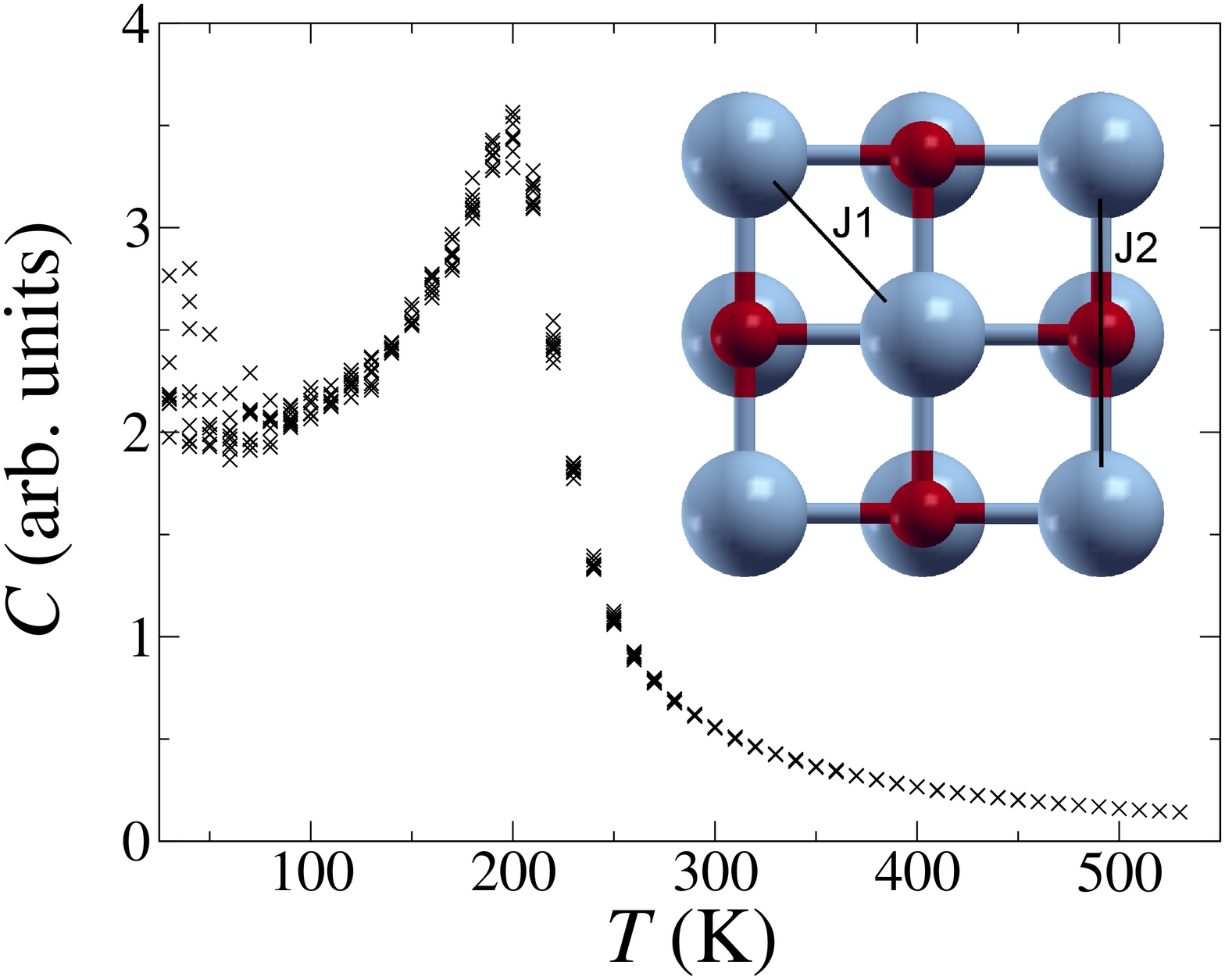}} 
\caption{Monte-carlo calculated specific heat for cubic CoO. Note a rather sharp peak at $T\sim210$~K,
that we associate with the N\'eel temperature $T_\mathrm{N}$. In the inset we present the RS CoO cell
and we indicate the various $J$ constants.}
\label{Fig3}
\end{figure}

The exchange constants $J_n$ are presented next (Tab. \ref{Tab1}). We find that the first and second
nearest neighbor constants $J_1$ and $J_2$ are sufficient to reproduce the DFT total energies with a
standard deviation of less than $3$~meV/Co. This corresponds to about 3\% of the total magnetic energy 
of the AFII structure (see Fig.~\ref{Fig2}).
The MC calculated specific heat, $C$, as a function of temperature, is presented in figure \ref{Fig3}. A clear
peak is observed, indicating that the N\'eel temperature is $T_\mathrm{N}\sim$~210~K. This also agrees with the value 
calculated by using the Binder cumulants and scaling theory, but it is 30\% lower than the
experimental $T_\mathrm{N}=$~287~K~\cite{CoOtn}. Considering the various approximations
introduced in our scheme, such as collinearity and the possible errors originating from the exchange and
correlation functional, we regard this value as satisfactory. Moreover, since 2+ oxidation state for Co is also observed 
for the WZ and ZB polymorphs, one can expect a similar underestimation of $T_\mathrm{C}$ ($\sim$30\%).

\section{Wurtzite cobalt oxide}
We now consider the WZ phase. Since this is the same lattice structure of ZnO, WZ CoO is the most likely
candidate as secondary phase in ZnO:Co. Although WZ CoO was first grown in the early sixties \cite{WUZnO62} 
and it can now be synthesized by several groups \cite{wurtzite_CoO2,liu_wurtzite_CoO,CoExp}, relatively little 
is known about its electronic and magnetic properties. Risbud et al. \cite{wurtzite_CoO2} found no 
ferromagnetism, but confirmed the presence of rather strong magnetic coupling between the Co ions. 
DFT-LDA calculations by the same authors indicate that the ferromagnetic state has lower energy than the 
non-magnetic one, although it is not necessarily the ground state. In fact a later study by Han et al. 
demonstrates that the ferromagnetic ground state is higher in energy than a magnetic configuration in which 
ferromagnetic $a,b$ planes align antiferromagnetically along the WZ $c$-axis ($c$-type antiferromagnetic) \cite{Han}. 
Whether or not this is  the ground state is unknown.

For the calculations of the WZ phase we use the experimental lattice parameters measured by Risbud~et al.~\cite{wurtzite_CoO2} 
($a$~=~3.244~\AA, $c$~=~5.203~\AA\ and $u$~=~0.416), which give a small pressure (18~kbar) and forces ($<$~0.005~eV/\AA).
The calculated paramagnetic energy per formula unit $E_0$ is 200.6~meV higher than that of the RS structure. This is obtained 
at a considerably larger volume (19.32~\AA$^3$ for RS, 23.71~\AA$^3$ for WZ), suggesting that the WZ polymorph will not 
form at equilibrium in the bulk. However, these are not large energy differences and one expects that the WZ phase can
be indeed stabilized in thin films or when alloying with ZnO. 

Interestingly the structure proposed by Risbud~et al.~\cite{wurtzite_CoO2} does not
appear to be a stable phase in DFT. Conjugate gradient relaxation move the oxygen atoms along the $c$-axis 
so that they lie in the same plane as Co ($u\rightarrow0.5$). For such a distorted phase (denoted as WZ$^*$)
$E_0$ is 80~meV lower than that of the experimental WZ phase and thus only 120~meV higher than that of the 
RS. WZ$^*$ has a volume slightly smaller than that of undistorted WZ, although still substantially larger than that 
of the naturally occurring RS. Interestingly the $c$-axis in WZ$^*$ is considerably compressed and the $a$ and $b$ 
axes are expanded with respect to the WZ phase. This distortion lowers the cell volume and increases the Co-O coordination 
number from 4 to 5. Since such a highly distorted phase has never been observed experimentally, we believe it may merely 
be an intermediary state in the transition from WZ to RS. This is supported by experimental evidence that the WZ polymorph 
is metastable and reverts back to the RS when annealed~\cite{liu_wurtzite_CoO}. Interestingly this 5-fold coordinated structure
has been previously predicted by DFT calculations for MgO~\cite{mgo}, when forced into a WZ crystal phase. Notably the
energy difference between the RS and the WZ phases in the case of CoO is considerably lower than in the case of MgO, 
explaining why WZ CoO has been synthesized, while WZ MgO has not. 

WZ CoO shows a similar electronic structure to that of RS with considerable O-$p$/Co-$d$ hybridization
in the valence band (Fig.~\ref{Fig1}). In this case the Mulliken populations of the Co $d$ orbitals
are found to be 4.76 and 2.12 for the majority and minority spins respectively. The oxygen atoms carry no 
magnetic moment so a total Mulliken magnetic moment of 2.64$\mu_\mathrm{B}$ is observed, consistent with 
the 2+ valence state. In the case of WZ CoO (both WZ and WZ$^*$) four $J$ constants are sufficient to yield a standard 
deviation of less than 1~meV/Co (see figure \ref{Fig2}).

\begin{figure}
\epsfxsize=8cm
\centerline{\epsffile{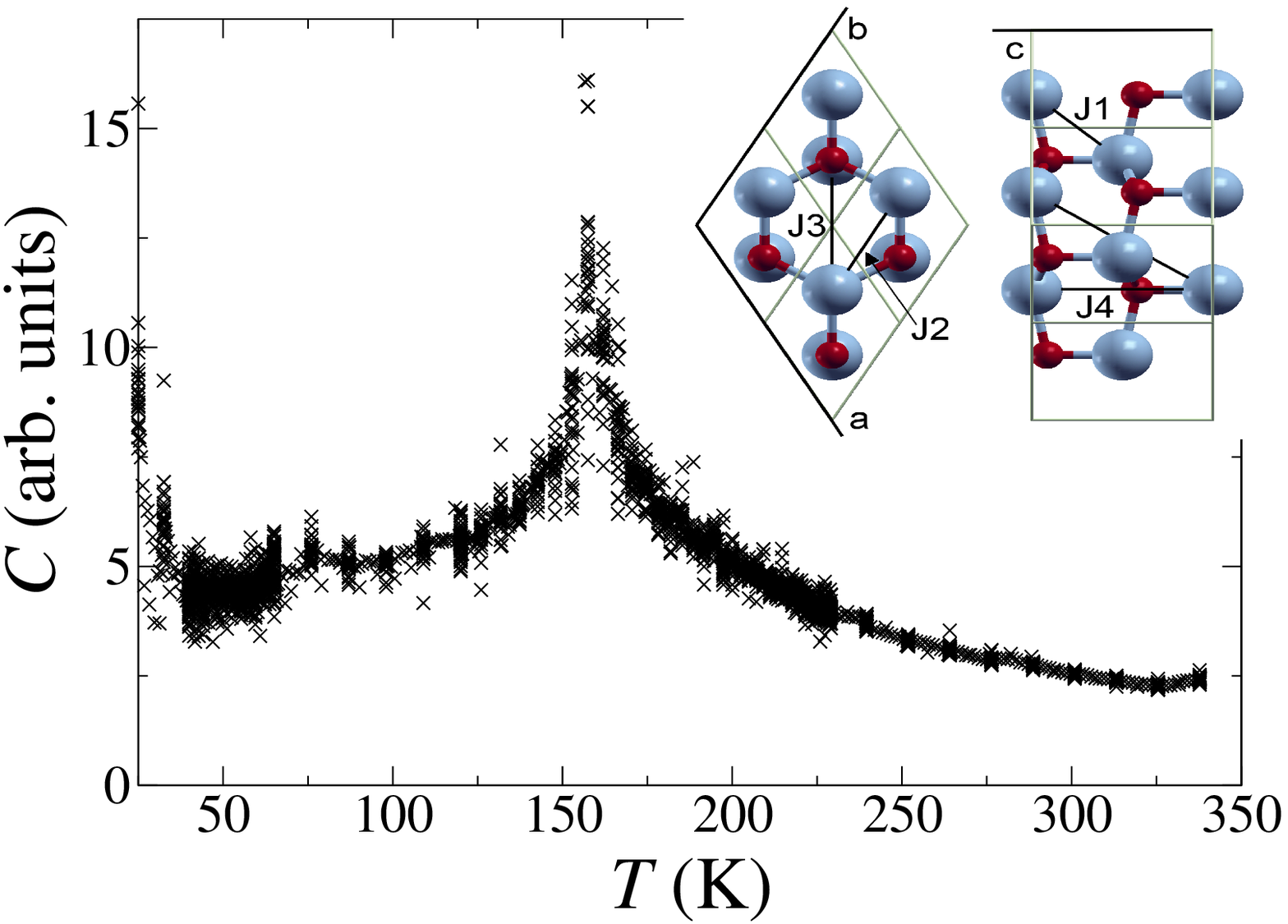}} 
\caption{Monte-carlo calculated specific heat for wurtzite CoO. Note a rather sharp peak at $T\sim160$~K,
that we associate with the magnetic critical temperature. In the inset we present the WZ CoO cell
and we indicate the various $J$ constants.}
\label{Fig4}
\end{figure}

The dominant interaction in WZ CoO is a strong nearest neighbor antiferromagnetic coupling in the
\{001\} planes, leading to a N\'eel frustrated state, in which adjacent spins in the planar triangular lattice
are rotated by 120$^\circ$ with respect to each other. Along the $c$ axis the interaction is ferromagnetic
between nearest neighbor planes and antiferromagnetic between second nearest neighbor, resulting in
an overall ferromagnetic coupling between the \{001\} planes. $C(T)$ for WZ CoO is presented in
figure \ref{Fig4}, from which we can estimate a critical temperature of about 160~K.
A similar analysis for the WZ$^*$ phase gives $T_\mathrm{C}~=~100$~K. Assuming that the error
found for the RS phase is transferable to WZ CoO, we obtain critical temperatures in the range 100-200~K.
These are well below room temperature and suggests that bulk WZ CoO cannot be
responsible for the room temperature magnetic signal often found in ZnO:Co.

\section{Zinc-blende Cobalt Oxide}

Finally we consider ZB CoO. This has been discovered experimentally during the synthesis of WZ
CoO, that indeed is always accompanied by the formation of the ZB phase~\cite{wurtzite_CoO2}. The 
ZB structure is now structurally characterized \cite{zincblende}, although no information is available
about its magnetic state. For the purpose of comparing the atomic and electronic structures we
set the spin configuration of the ZB cell to have ferromagnetic planes arranged in an antiferromagnetic
stack along the [100] direction. Our relaxed structure has a lattice parameter of 3.245~\AA, which compares well
with the experimental value of 3.230~\AA\ \cite{zincblende}. Similarly to the other polymorphs, ZB CoO shows a 
strong O-$p$/Co-$d$ hybridization in the valence band (Fig.~\ref{Fig1}) and the Mulliken magnetic
moment is around 3~$\mu_\mathrm{B}$ (2.74~$\mu_\mathrm{B}$ with a Co $d$ Mulliken occupation of
4.77 and 2.03 for the majority and minority spins respectively).
\begin{figure}
\epsfxsize=8cm
\centerline{\epsffile{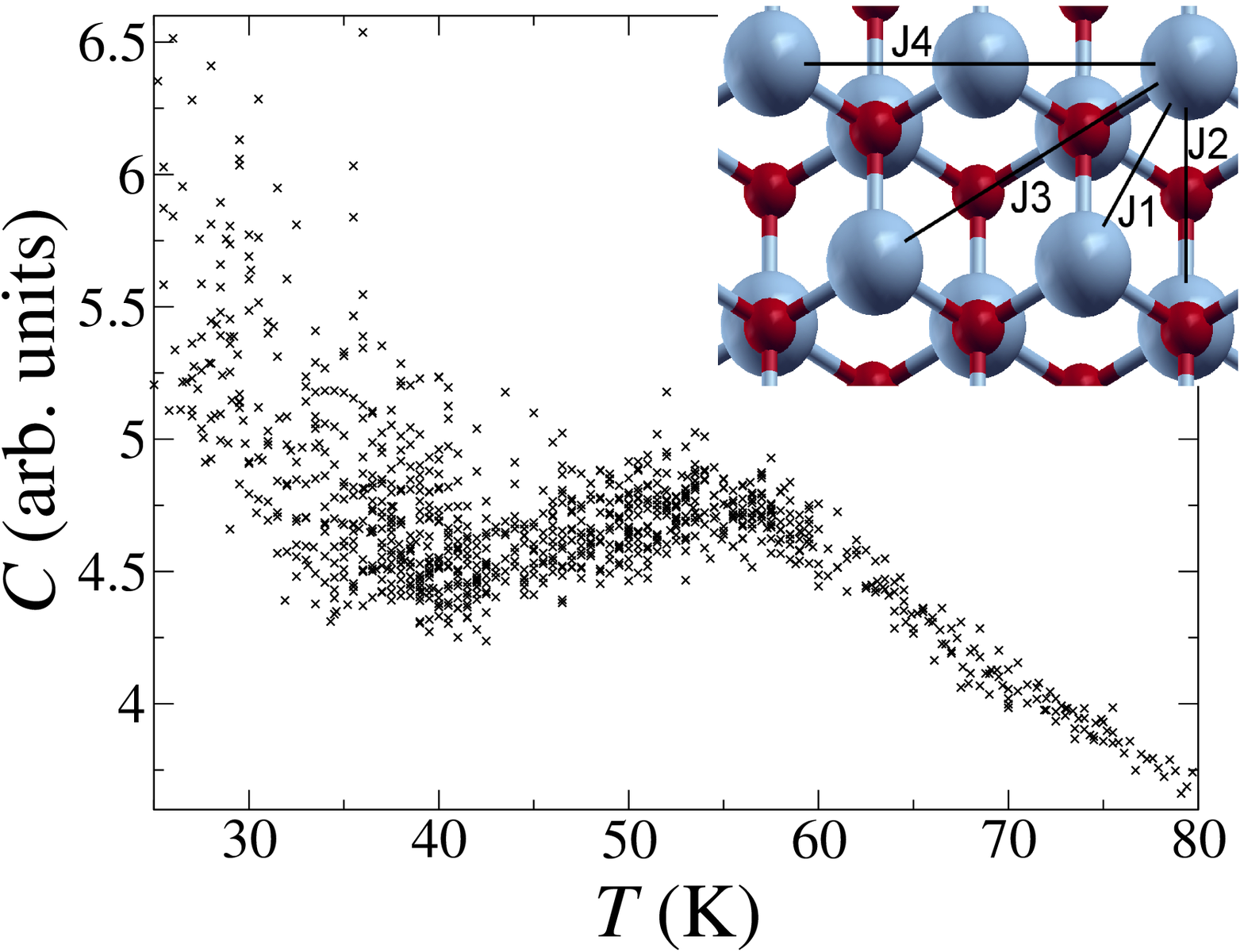}} 
\caption{Monte-carlo calculated specific heat for zincblende CoO. Note a rather diffuse peak at $T\sim55$~K,
that we associate with the magnetic critical temperature. In the inset we present the ZB CoO cell
and we indicate the various $J$ constants.}
\label{Fig5}
\end{figure}
The ZB phase is found to be the least stable phase among all the CoO polymorphs studied, with $E_0=313.2$~meV. 
At equilibrium the volume is essentially identical to that of WZ and the energy difference between the ZB and 
WZ phases is small, supporting the experimentally observed co-existence of the WZ and ZB phases~\cite{wurtzite_CoO2}.

The first four $J$s already describe accurately the total magnetic energy of the ZB phase
with a standard deviation of less than 1~meV/Co. We find that in the case of ZB CoO the first
nearest neighbor interaction is by far the largest and accounts for most of the magnetic energy. This however is
considerably lower than the dominant $J$ for both the WZ and the RS phases and one expects a considerably
lower critical temperature. This is confirmed by our MC simulations (Fig.~\ref{Fig5}), which gives
us $T_\mathrm{C}=$~55~K. Importantly $J_1$ is antiferromagnetic leading to three-dimensional
frustration, evident in the rather diffuse peak in $C(T)$.

\section{Conclusions}

In conclusion a combination of DFT and Monte Carlo methods have been used to calculate the thermodynamic
properties of CoO polymorphs in their bulk phases. The scheme was tested first for the RS phase and then applied to both the WZ 
and the ZB structures. Interestingly the ground state of the three polymorphs is rather different. RS CoO has a type-II 
antiferromagnetic structure, WZ CoO is a two-dimensional frustrated system and ZB CoO is three dimensionally 
frustrated. In addition a second WZ structure was identified. Crucially, despite these differences, all the polymorphs 
show critical temperature considerably below room temperature. Although our results are for the bulk phases we would 
expect them to be applicable to the central atoms of a cluster. Therefore bulk CoO clusters in any crystalline phase 
cannot be the explantion for the experimentally observed room temperature magnetism of diluted ZnO:Co. Whether this
can originate from uncompensated spins at the surface of those clusters however cannot be ruled out.  \\
\\
This work is sponsored by Science Foundation of Ireland (07/IN.1/I945). We thank TCHPC
and ICHEC for providing computational support.

\bibliographystyle{plain}

\end{document}